\documentclass[graybox,natbib,nosecnum]{svmult}
\bibpunct{(}{)}{;}{a}{}{,} 

\pdfoutput=1   

\usepackage{mathptmx}       
\usepackage{helvet}         
\usepackage{courier}        
\usepackage{type1cm}        

\usepackage{makeidx}         
\usepackage{graphicx}        
\usepackage{multicol}        
\usepackage[bottom]{footmisc}
\usepackage[normalem]{ulem}	

\usepackage{textcomp}

\usepackage{hyperref}  

\usepackage{amssymb,amsmath} 
\newcommand{\numax}{\mbox{$\nu_{\rm max}$}}
\newcommand{\dnu}{\mbox{$\Delta \nu$}}

\newcommand{\kp}{\mbox{\textit{Kepler}}}
\newcommand{\teff}{\mbox{$T_{\rm eff}$}}
\newcommand{\logg}{\mbox{$\log g$}}
\newcommand{\feh}{\mbox{$\rm{[Fe/H]}$}}

\newcommand{\rsun}{\mbox{$R_{\odot}$}}

\begin{document}

\title*{Using asteroseismology to characterise exoplanet host stars
}

\author{Mia S. Lundkvist, Daniel Huber, Victor Silva Aguirre, and William J. Chaplin}

\institute{
Mia S. Lundkvist \at Zentrum f{\"u}r Astronomie der Universit{\"a}t Heidelberg, Landessternwarte, K{\"o}nigstuhl 12, 69117 Heidelberg, DE, and Stellar Astrophysics Centre, Aarhus University, Ny Munkegade 120, 8000 Aarhus C, DK,
\email{lundkvist@phys.au.dk}
\and Daniel Huber \at Institute for Astronomy, University of Hawai`i, 2680 Woodlawn Drive, Honolulu, HI 96822, US and Sydney Institute for Astronomy, School of Physics, University of Sydney, NSW 2006, Australia,
\email{huberd@hawaii.edu}
\and Victor Silva Aguirre \at Stellar Astrophysics Centre, Aarhus University, Ny Munkegade 120, 8000 Aarhus C, DK,
\email{victor@phys.au.dk}
\and William J. Chaplin \at School of Physics and Astronomy, University of Birmingham, Edgbaston, Birmingham B15 2TT,
UK, \email{w.j.chaplin@bham.ac.uk}}

\maketitle

\abstract{
The last decade has seen a revolution in the field of asteroseismology -- the study of stellar pulsations. It has become a powerful method to precisely characterise exoplanet host stars, and as a consequence also the exoplanets themselves. This synergy between asteroseismology and exoplanet science has flourished in large part due to space missions such as \kp, which have provided high-quality data that can be used for both types of studies. 
Perhaps the primary contribution from asteroseismology to the research on transiting exoplanets is the determination of very precise stellar radii that translate into precise planetary radii, but asteroseismology has also proven useful in constraining eccentricities of exoplanets as well as the dynamical architecture of planetary systems.
In this chapter, we introduce some basic principles of asteroseismology and review current synergies between the two fields.
}

\section{Introduction}
"You only know your planet as well as you know your star". This could be the beginning of a sales pitch for asteroseismology -- the study of stellar pulsations -- as an important tool for exoplanet scientists to gain valuable information about their planets and planetary systems. This is because one of the strengths of asteroseismology is the ability to determine very precise stellar radii that can in turn be used to calculate the planetary size which, from the transit light curve, is known only as a function of the stellar size. Asteroseismology can also yield, for instance, the stellar age and luminosity, parameters that are important in order to assess the habitability of other planets.

Over the last decade, the field of asteroseismology has been revolutionised by space-based photometry provided by missions such as CoRoT and \kp. Before the success of these missions, asteroseimic studies of even a single star were time-consuming and had only been carried out for a few targets \citep[e.g. $\alpha$~Cen~A,][]{ref:bouchy2001,ref:bedding2004}. Today, oscillations have been detected in over 500 solar-type stars \citep{ref:chaplin2014}, a significant fraction of which are exoplanet host stars.

In this chapter we will introduce the basic principles of asteroseismology, describe how stellar properties can be derived from this and discuss the resulting precision. Hereafter we will focus on the synergy between asteroseismology and exoplanet science by highlighting how the fields combine and benefit from each other through some specific examples \citep[see also][]{ref:huber2018}.

\section{Introduction to asteroseismology}
Many different types of stars have been found to oscillate due to standing waves in their interiors. These oscillations give rise to detectable periodic changes in the brightness of the stars, and asteroseismology is the study of these oscillations. In this chapter we focus on solar-like oscillations, which are stochastically excited and damped by near-surface convection and have been observed in both solar-type stars and red giants. Several excellent reviews already exist on this topic and for further details the reader is referred to, for instance, \citet{ref:aerts2010}, \citet{ref:chaplin2013} or \citet{ref:bedding2014}. Below we will give a brief overview of important aspects of solar-like oscillations.

\subsection{Description of oscillation modes}
There are two main types of oscillation modes in solar-type stars: \textit{p-modes}, for which the restoring force is the pressure gradient, and \textit{g-modes}, for which the restoring force is buoyancy. Some oscillations can also exhibit a mixed character (\textit{mixed modes}), displaying p-mode-like behaviour in the stellar envelope and g-mode-like behaviour closer to the core.

The modes can be described mathematically using three integers. The first of these is the radial order ($n$), which is related to the number of radial nodes (nodal shells) of the standing wave inside the star. The radial order $n$ is positive for p-modes while it is negative for g-modes. The angular degree ($\ell$) specifies the number of nodal lines on the stellar surface. Due to cancellation effects when observing a non-spatially resolved star, only modes of low angular degree can typically be observed ($\ell \lesssim 4$). Modes with $\ell = 0$ are radial modes, while modes with $\ell \geq 1$ are non-radial modes. The radial oscillation modes are the simplest modes in which a star can pulsate, since the star heats and cools, contracts and expands, in a spherically symmetrical fashion.

The last integer needed to specify a mode is the azimuthal order ($m$), whose absolute value gives the number of surface nodes that cross the stellar equator. The value of $m$ can be any integer in the range from $-\ell$ to $+\ell$, thus there are $2\ell + 1$ possible azimuthal orders per degree. However, the oscillation frequencies are independent of the $m$-value unless spherical symmetry is broken, which, for example, is the case if the star is rotating. In this case, each oscillation mode with a given $n$ and $\ell$ value will split in frequency and reveal the different $m$-components. The frequency splitting between each $m$-component is governed by the rotation in those interior layers of the star to which the mode is sensitive. Since energy equipartition holds to a very good approximation for moderately rotating stars, their relative amplitudes are set by the inclination angle of the stellar rotation axis with respect to the line of sight. As a consequence, information about the rotation axis and speed of the star can be extracted from the observed splitting of these multiplets. This will be discussed further later in this chapter.

\subsection{The frequency spectrum}
\begin{figure}
	\centering
	\includegraphics[scale=.38]{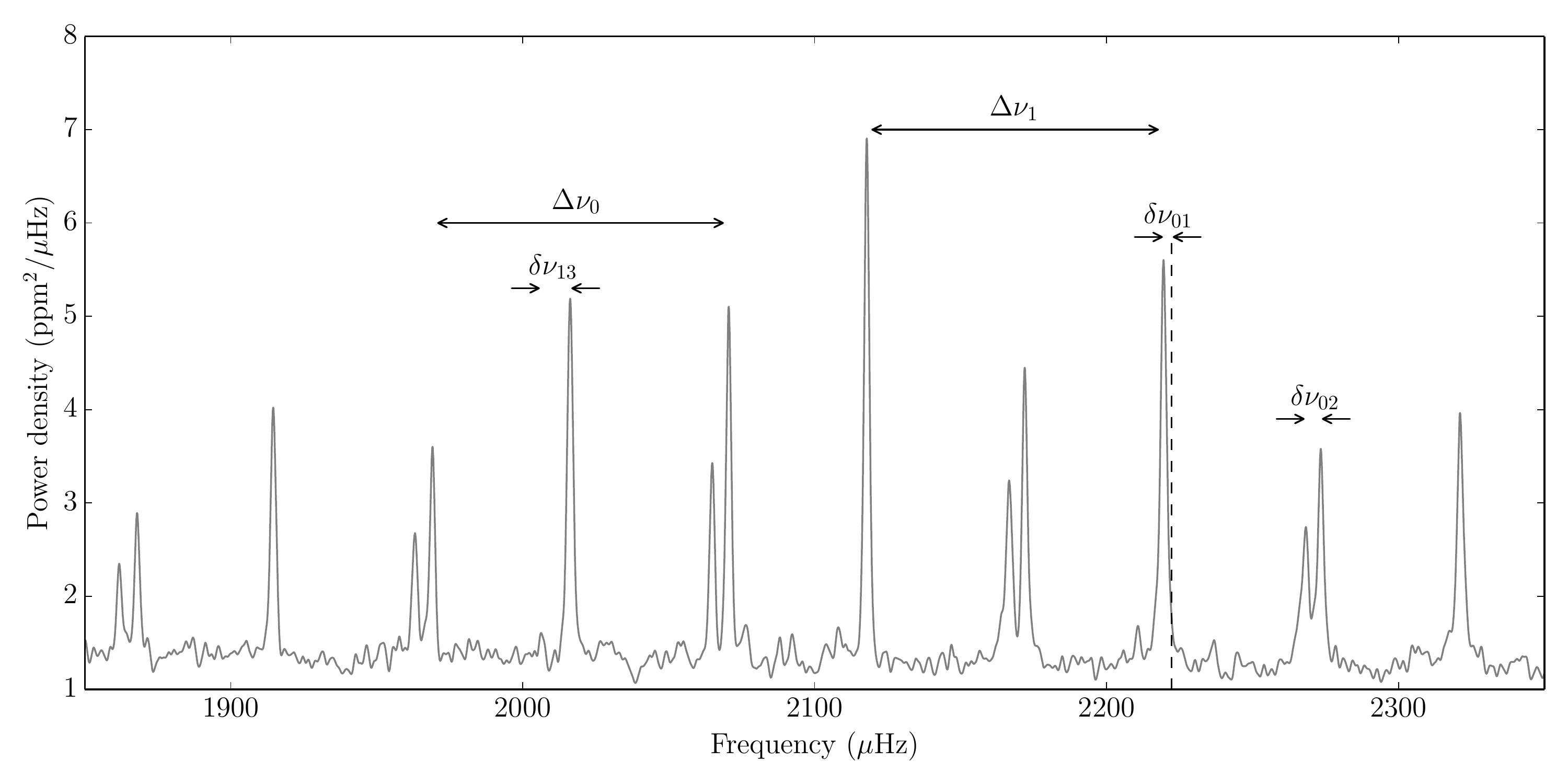}
	\caption{Power spectrum of the solar-like oscillator Kepler-68 (KIC 11295426) from \kp\ data. A segment of the power spectrum is shown, centered on the region of largest oscillation amplitudes, clearly revealing the regular p-mode structure. The large and small separations are indicated. The spectrum has been smoothed twice with a Gaussian filter with a width of $1 \ \mu$Hz.}
	\label{fig:pwr_spec}
\end{figure}
The frequency spectrum of a typical solar-type star observed by \kp, Kepler-68 (KIC 11295426), is shown in Fig.~\ref{fig:pwr_spec}. It depicts the power as a function of frequency with the radial order increasing with frequency, while the angular degree follows a regular pattern of alternating spherical degrees.
The peaks with the largest amplitudes are located in the middle of the excited frequency range, and the centre of the power envelope defines the \textit{frequency of maximum oscillations power} (\numax).

The p-modes in a solar-like oscillator follow a regular structure. Dominating this regularity is the \textit{large frequency separation} (\dnu), which is the average frequency spacing between modes with consecutive radial orders and the same angular degree. Observationally and theoretically \dnu\ is most commonly defined by the average separation of radial modes, since these frequencies are unaffected by frequency shifts due to mixed modes (since no radial g-modes exist).

The small frequency separations $\delta\nu_{02}$ and $\delta\nu_{13}$ are those between modes with a radial order that differs by unity and an angular degree that differs by two. These separations have also been indicated in Fig.~\ref{fig:pwr_spec} in addition to another small frequency separation, namely the offset of the $\ell = 1$ peaks from the midway point between two modes with $\ell = 0$ ($\delta\nu_{01}$). Deviations from regularity in the large and small frequency separations are due to rapid changes in the sound speed profile (so-called acoustic glitches), and hence are powerful diagnostics for infer the interior structure of stars.

Frequencies of high radial order and low angular degree can be described using asymptotic theory \citep{ref:tassoul1980}. In this framework, the large frequency separation can be shown to be given by the inverse of the sound travel time across the stellar diameter \citep[e.g.][]{ref:aerts2010}:
\begin{equation}
	\label{eq:cs_int}
    \Delta\nu = \left( 2 \int\limits_0^R \frac{\mathrm{d}r}{c} \right) ^{-1} \ .
\end{equation}
Here, $c$ is the sound speed and $R$ the stellar radius. The large frequency separation has been found to scale to a good approximation with the mean stellar density, while the small frequency separations are sensitive to the stellar age since they depend on the sound speed gradient in the deep stellar interior. This means that the frequency separations of a star of given mass and composition will vary in time as a result of stellar evolution.
%
\section{Stellar properties from scaling relations}
Scaling relations based on the average large frequency separation and the frequency of maximum oscillation power are a powerful way to determine fundamental stellar properties for a large number of stars. This approach can be used when the signal-to-noise ratio in the power spectrum is too low to allow for a reliable extraction of individual oscillation frequencies. In this section the scaling relations will be derived and discussed in terms of their accuracy and suggested modifications.
%
\subsection{Scaling relations for \texorpdfstring{\dnu}{the large frequency separation} and \texorpdfstring{\numax}{the frequency of maximum power}}
Assuming adiabatic conditions and an ideal gas we can express the sound speed $c$ and temperature $T$ as $c \propto\sqrt{T/\mu}$ and $T\propto \mu M / R$, where $\mu$ is the mean molecular weight. Substituting these expressions into Eq.~\ref{eq:cs_int}, it can be seen that \dnu\ is proportional to the square root of the mean stellar density \citep{ref:ulrich1986,ref:kjeldsen1995,ref:bedding2010}:

\begin{equation}
	\label{eq:dnu_sca}
    \Delta\nu \propto \sqrt{\bar{\rho}} \propto \sqrt{\frac{M}{R^3}} \ ,
\end{equation}
with $M$ being the stellar mass. This implies that as a solar-type star expands during its evolution along the main sequence and beyond, the mean density and thus the large frequency separation will decrease.

The frequency of maximum power (\numax) is another observable that changes as a star evolves. This can be understood through its suggested scaling with the acoustic cut-off frequency \citep[see for instance][]{ref:brown1991,ref:kjeldsen1995,ref:belkacem2011}. The acoustic cut-off frequency for an isothermal atmosphere is given by
\begin{equation}
	\label{eq:acousticcutoff}
	\nu_\mathrm{ac} = \left( \frac{c}{4\pi H} \right) \ ,
\end{equation}
$H = - \left( \mathrm{d}\ln \rho / \mathrm{d}r \right)^{-1}$ being the density scale height, which is proportional to $g T_\mathrm{eff}^{-1/2}$ with $g$ being the surface gravity and $T_\mathrm{eff}$ the effective temperature. Thereby we have \citep{ref:kjeldsen1995}
\begin{equation}
	\label{eq:numax_sc}
    \nu_\mathrm{max} \propto \frac{g}{\sqrt{T_\mathrm{eff}}}
    				 \propto \frac{M}{R^2 \sqrt{T_\mathrm{eff}}} \ .
\end{equation}
Thus, as above, when a solar-type star evolves, the frequency of maximum power decreases due to a decreasing surface gravity.

The scaling relations for \dnu\ and \numax\ (Eqs.~\ref{eq:dnu_sca} and~\ref{eq:numax_sc}) can be combined with the stellar effective temperature to yield fundamental stellar properties such as mass and radius. It is worth noting, however, that these relations are extrapolated from the well known solar properties. Thus, an inherent assumption when using the scaling relations for stars ranging from the main sequence to red giant phases is that they are homologous to the Sun. In this manner, the following equations for mass, radius, surface gravity, and mean density are obtained:

\begin{equation}
	\label{eq:m_direct}
	\frac{M}{M_\odot} \cong
			\left( \frac{\nu_\mathrm{max}}{\nu_{\mathrm{max},\odot}} \right) ^3 				
			\left(\frac{\Delta\nu}{\Delta\nu_\odot}\right) ^{-4}
			\left( \frac{T_\mathrm{eff}}{\mathrm{T}_{\mathrm{eff},\odot}} \right) ^{1.5} \ ,
\end{equation}
\begin{equation}
	\label{eq:r_direct}
	\frac{R}{R_\odot} \cong
			\left( \frac{\nu_\mathrm{max}}{\nu_{\mathrm{max},\odot}} \right) 
			\left(\frac{\Delta\nu}{\Delta\nu_\odot}\right) ^{-2}
			\left( \frac{T_\mathrm{eff}}{\mathrm{T}_{\mathrm{eff},\odot}} \right) ^{0.5} \ ,
\end{equation}
\begin{equation}
	\label{eq:g_direct}
	\frac{g}{g_\odot} \cong
			\left( \frac{\nu_\mathrm{max}}{\nu_{\mathrm{max},\odot}} \right)
			\left( \frac{T_\mathrm{eff}}{\mathrm{T}_{\mathrm{eff},\odot}} \right) ^{0.5} \ ,
\end{equation}
\begin{equation}
	\label{eq:rho_direct}
	\frac{\bar{\rho}}{\bar{\rho}_\odot} \cong
			\left(\frac{\Delta\nu}{\Delta\nu_\odot}\right) ^{2} \ .
\end{equation}
This is sometimes referred to as the direct method to obtain fundamental properties, and it provides estimates that are independent of stellar evolutionary models under the assumption that the scaling relations are valid \citep[see for instance][]{ref:chaplin2013}. It can be seen from Eqs.~\ref{eq:m_direct} and~\ref{eq:r_direct} that the determination of stellar mass is inherently more uncertain than that of the radius, since it depends more strongly on the parameters in the expression. The equations can also be written in several other ways, for instance using the relation between radius, temperature and luminosity; $L \propto R^2 T_\mathrm{eff}^4$ \citep[see, for example,][]{ref:bedding2014}.\newline

A disadvantage of the direct method is that the propagated uncertainties for masses and radii can be significantly overestimated since our physical knowledge of stellar evolution is ignored (i.e., any combination of $T_\mathrm{eff}$, $R$ and $M$ are allowed). 
Another disadvantage of the direct method is that it cannot estimate ages. These issues can be circumvented by instead adopting a grid-based modelling approach, where we compare the observed \dnu\ and \numax\ to predictions from stellar evolutionary models. The "best matching" models are then taken as having properties that closely match those of the target star. The approach for \dnu\ is straightforward: one applies a linear fit as a function of radial order to the theoretically computed frequencies of the stellar model weighted in a manner that mimics the analysis used to extract the observed \dnu\ from the real data \citep{white11}. Unfortunately, matters are less straightforward for \numax\ owing to shortcomings in non-adiabatic predictions of the excitation and damping of the modes (which are required to compute a model \numax). Instead, a common approach is to use the mass, radius and temperature of the model as input to the \numax\ scaling relation to compute the model-predicted parameter.

\subsection{Accuracy of the scaling relations}

Equations \ref{eq:dnu_sca} and \ref{eq:numax_sc} are approximate relations which require careful validation. Due to the wealth of space-based data and the complexity of modeling oscillation frequencies for large numbers of stars, many studies have relied on scaling relations and hence testing their accuracy has become one of the most active topics of research in asteroseismology. 

\begin{figure}
\begin{center}
\resizebox{\hsize}{!}{\includegraphics{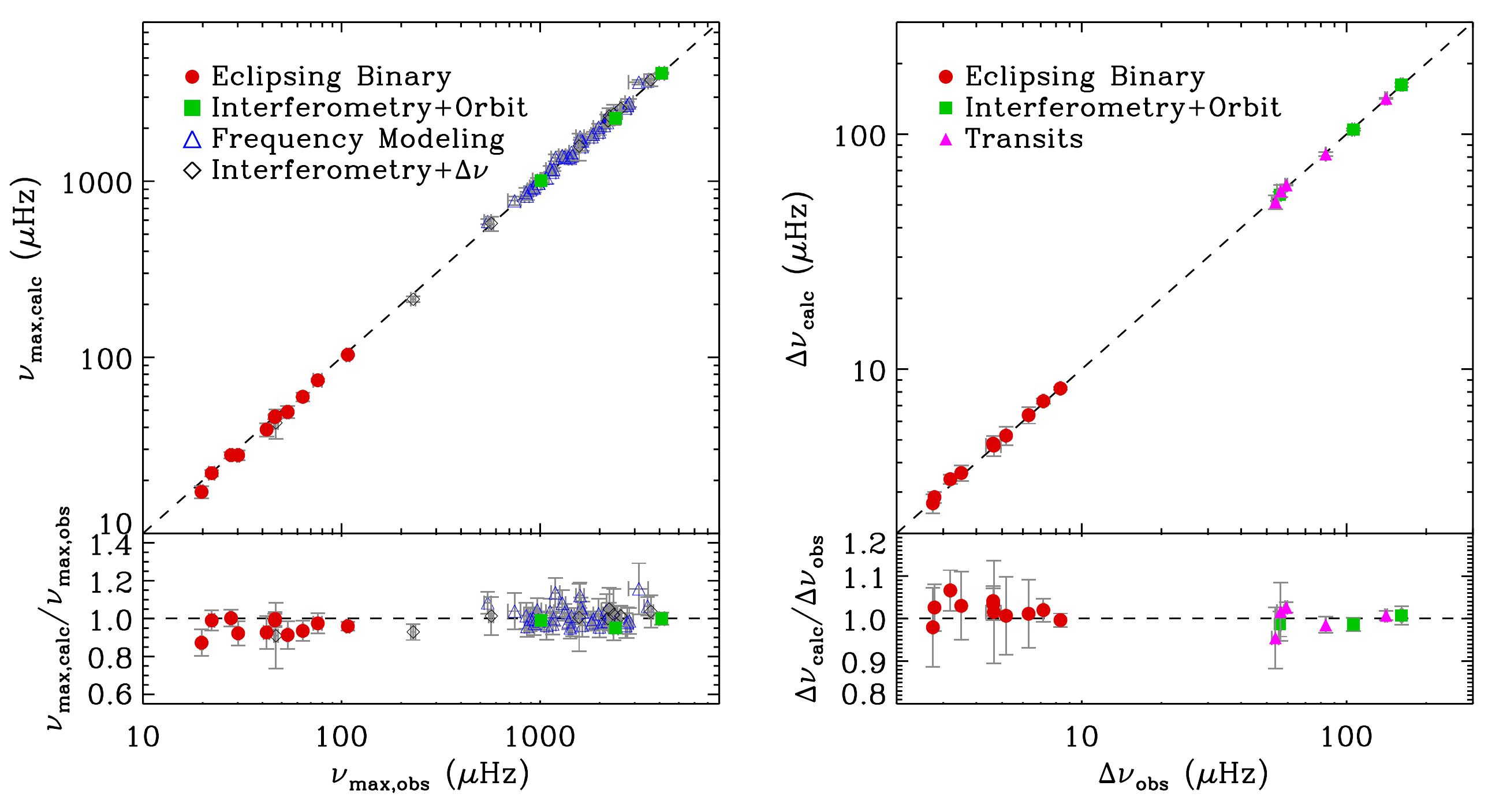}}
\caption{Empirical tests of asteroseismic scaling relations. Left: Comparison of measured \numax\ values to values calculated from independent measurements (see legend). Filled symbols are empirical values which are independent of asteroseismology. Right: Same as left panel but for the \dnu\ scaling relation. Adapted from \citet{huber14b}.}
\label{fig:scalingtest1}
\end{center}
\end{figure} 
 
\noindent
\textbf{Density and Surface Gravity:} Direct tests of Equations \ref{eq:dnu_sca} and \ref{eq:numax_sc} require independent measurements of \logg\ or density, which are typically only possible for eclipsing, spectroscopic binaries or the combination of interferometry with astrometric orbits. Figure \ref{fig:scalingtest1} shows a comparison of measured and calculated \numax\ and \dnu\ values for systems where such constraints are currently available. The comparison also includes densities derived from transiting exoplanets with independently constrained eccentricities such as HD17156 \citep{nutzman11,gilliland11}, TrES-2 \citep{southworth11,barclay12}, HAT-P-7 \citep{cd10,southworth11}, Kepler-14 \citep{southworth12,ref:huber2013}, and Kepler-7 \citep{ref:sanford2017}. Furthermore, tests of \numax\ through more indirect methods include combining interferometric angular diameters with densities calculated from the \dnu\ scaling relation, or masses and radii determined from individual frequency modeling \citep[e.g.,][]{metcalfe14}. Figure \ref{fig:scalingtest1} demonstrates good agreement over three orders of magnitude, with median residuals indicating typical accuracies of 0.03\,dex and 0.01\,dex in \logg\ for giants and dwarfs/subgiants, as well as 4\% and 2\% in density for giants and dwarfs/subgiants.

\noindent
\textbf{Stellar Masses:} Figure \ref{fig:scalingtest1} shows evidence that deviations increase for more evolved stars, as expected since scaling relations are anchored to the Sun. There has been mounting evidence that asteroseismic masses for red giants are overestimated by $\sim5-15\%$ based on comparisons with dynamical masses from eclipsing binaries \citep{frandsen13,gaulme16}, cluster masses determined from near turn-off eclipsing binaries \citep{brogaard12}, and expectation values for luminous metal-poor ($\feh < -1$) giants \citep{epstein14}. While the extent and dependence of this offset on evolutionary state is not yet clear, it indicates that masses for red giants from scaling relations should be broadly accurate to $\approx$\,10\%. Independent masses of seismic dwarfs and subgiants are rare, but can be expected to better than 5\%.

\begin{figure}
\begin{center}
\resizebox{10cm}{!}{\includegraphics{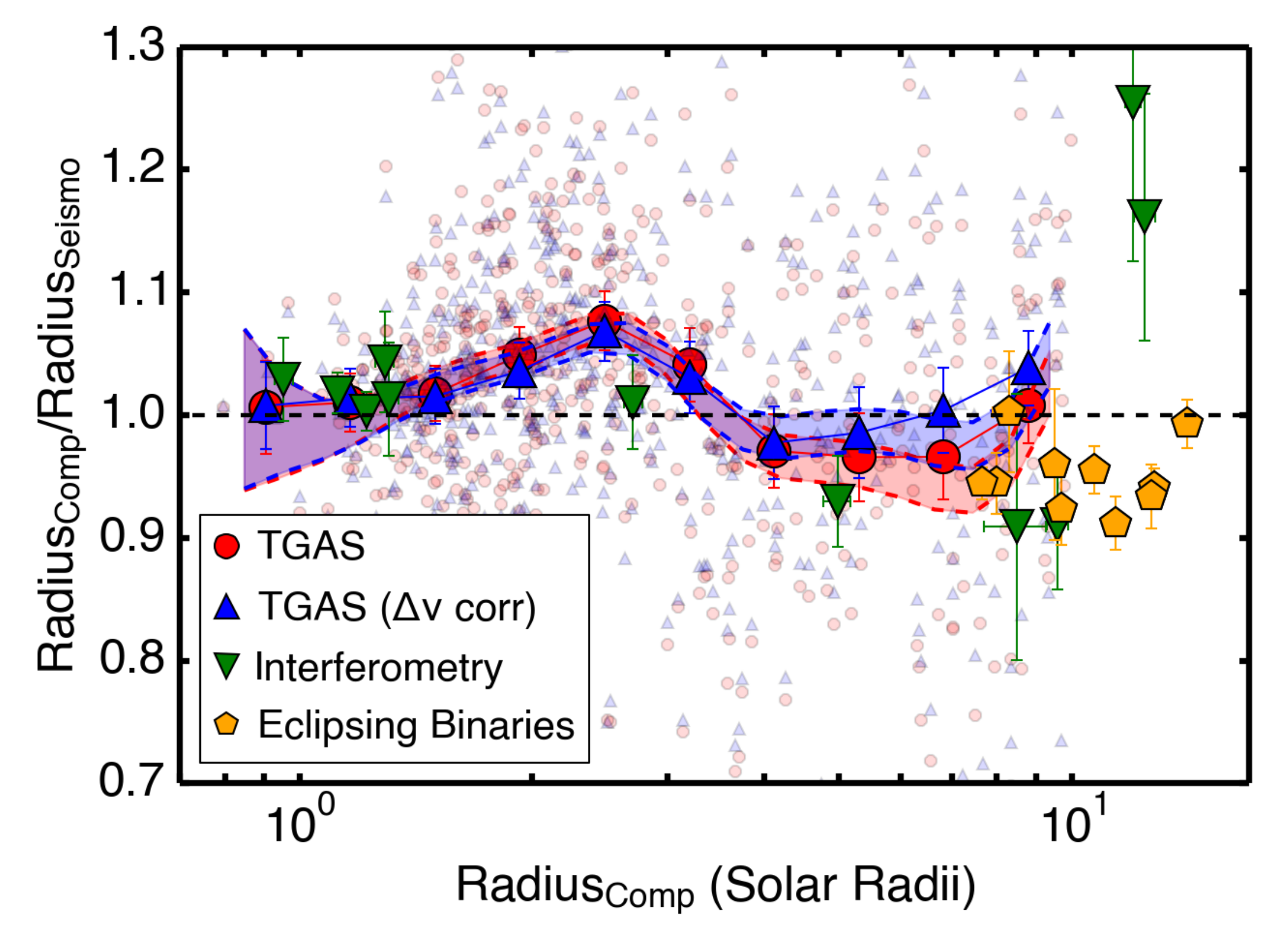}}
\caption{Comparison of asteroseismic radii derived from scaling relations with radii derived from four methods. Red circles and blue upward triangles show the TGAS (Tycho-Gaia Astrometric Solution) sample with and without the \citet{sharma16} \dnu\ scaling relation correction (small symbols show unbinned data), and shaded areas show 68\% confidence intervals. We also show stars with interferometrically measured radii  \citep[green triangles,][]{huber12,white13,johnson14} and red giants in double-lined eclipsing binary systems \citep[orange pentagons,][]{gaulme16}. From \citet{huber17}, \copyright AAS. Reproduced with permission.}
\label{fig:scalingtest2}
\end{center}
\end{figure}

\noindent
\textbf{Stellar Radii:} Tests of stellar radii typically result from the combination of angular diameters measured through optical long-baseline interferometry with parallaxes \citep{huber12b,white13,johnson14}, followed by more indirect estimates from luminosities derived from parallaxes \citep{silva12,huber17} or clusters \citep{miglio11, miglio16}. Figure \ref{fig:scalingtest2} shows a summary of current empirical tests of radii from scaling relations. While no significant systematic offsets are evident for main-sequence stars ($\lesssim 1.5 \ \rsun$), comparisons to Gaia radii and eclipsing binaries show evidence for systematic differences at the $\approx$\,5\% level for subgiants ($\approx$1.5--3 \ \rsun) and evolved red giants ($>$8 \ \rsun). Overall, these tests indicate that scaling relation radii are accurate to a few percent for dwarfs/subgiants and to within $\approx$\,5\% for giants.

\noindent
\textbf{Scaling Relation Corrections:} Due to the lack of empirical results spanning a large range of parameter space, corrections to scaling relations have so far mostly relied on theoretical work. Specifically, most proposed corrections have compared the average large frequency separation (\dnu) calculated from individual frequencies with model densities \citep{stello09c,white11,guggenberger16,sharma16} or suggested an extension of the asymptotic relation \citep{mosser13}. A consistent result is that $\dnu$ scaling relation corrections should depend on \teff, evolutionary state and metallicity, and first results have shown that such corrections indeed improve the agreement with independent constraints \citep{sharma16,huber17}. Uncertainties in modeling the driving and damping of oscillations typically prevent theoretical tests of the $\numax$ scaling relation, although some studies have shown encouraging results \citep{ref:belkacem2011}.

\section{Stellar properties from individual frequencies}

Instead of the global asteroseismic parameters, the individual oscillation frequencies can be used to infer stellar properties. The increase in the information content provided by the asteroseismic data leads to a higher precision than what can be achieved using \dnu\ and \numax.

In this approach, the individual oscillation frequencies are compared to frequencies determined from stellar models in order to obtain the best possible agreement. In addition to the observed frequencies, the stellar effective temperature and metallicity are used as external observables to place extra constraints on the models. However, because of improper modelling of the near-surface layers, there is an offset between observed and computed oscillation frequencies, the so-called surface term or surface effect. Commonly, the offset is corrected using empirical or theoretically motivated schemes such as those proposed by \citet{ref:kjeldsen2008,ref:ball2014,ref:sonoi2015}. It is possible to avoid the need for a correction if one fits frequency differences or frequency  difference ratios instead, since these are combinations of frequencies that minimise the impact of the near-surface layers \citep[e.g.,][]{roxburgh03,SilvaAguirre:2011jz}.

\begin{figure}[htbp]
  \centering
    \includegraphics[width=.8\textwidth]{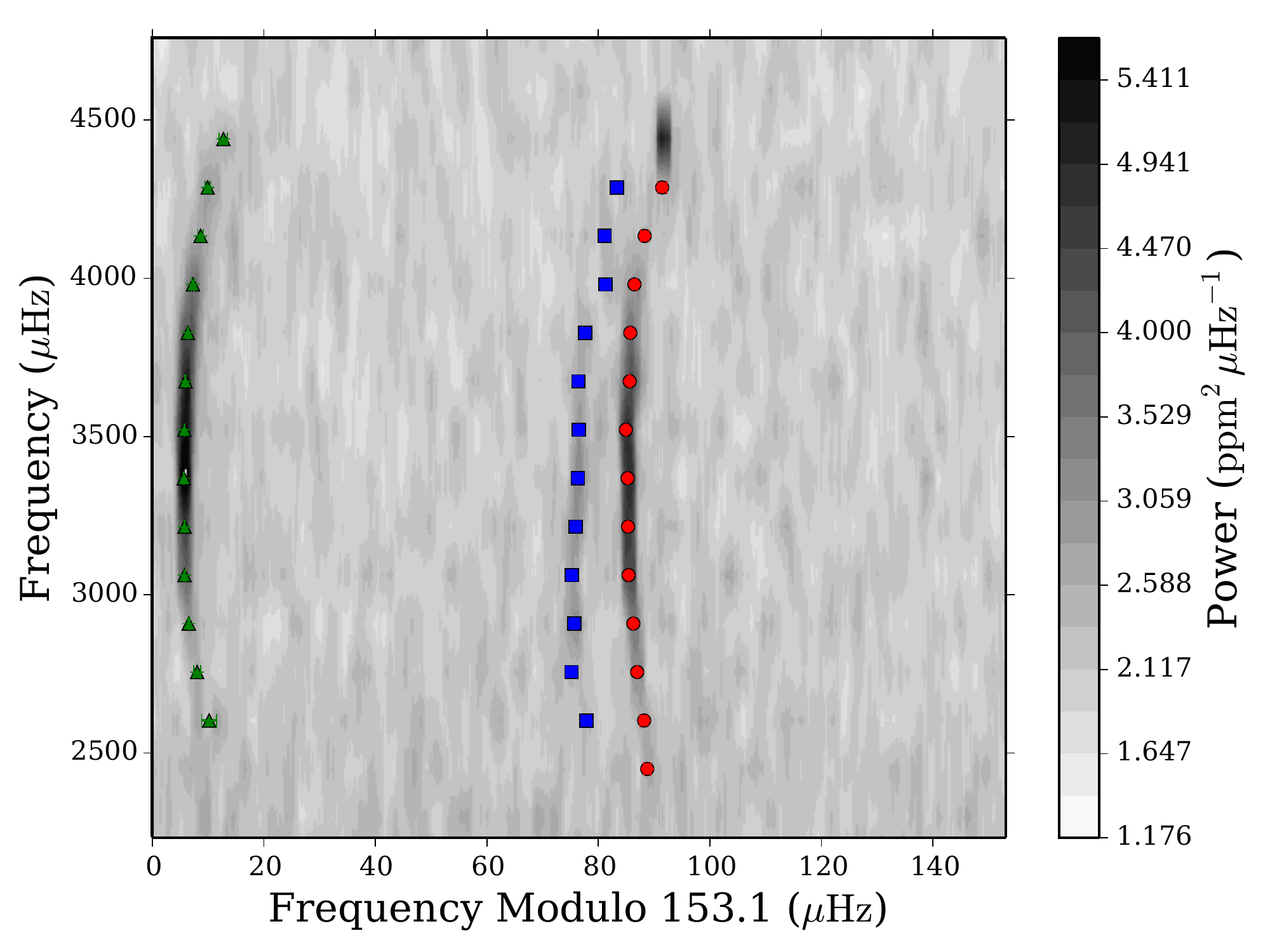}
\caption{{\'E}chelle diagram of Kepler-409 showing the power in greyscale and the model-frequencies in colour on top. The modes are labelled according to angular degree with the red circles showing the radial modes ($\ell = 0$), the green triangles giving the dipolar modes ($\ell = 1$) and the blue squares indicating the quadrupolar modes ($\ell = 2$). From \citet{ref:silvaaguirre2015}.}
\label{fig:echelle}
\end{figure}

An example of a best-fitting model, i.e., a model that is able to reproduce the observed frequency spectrum, can be seen in Fig.~\ref{fig:echelle}. This \textit{{\'e}chelle diagram} \citep{ref:grec1983} (or ladder diagram) is often used to display modelled and observed oscillation frequencies. It can be constructed by dividing the power spectrum into segments of length \dnu\ in frequency, and then stacking them on top of each other. The oscillation modes with the same angular degree will line up vertically, forming a separate ridge for each degree, with mixed modes showing a strong deviation from the ridge structure. If the frequency spacings did not vary with frequency, each ridge would be perfectly straight (and vertical if the correct \dnu\ is used), but since the spacings do change slightly with frequency, the ridges have some curvature.

In addition to constraining the fundamental stellar properties, including the age of the star, modelling the individual frequencies also allows determination of the surface helium abundance and the depth of the outer convection zone \citep[e.g.,][]{Verma:2014fx,Mazumdar:2014kt}. Not only can more parameters be constrained using detailed modelling of the observed frequencies, the precision on the fundamental properties is also improved compared to the global-parameters case. For $66$ of the best-quality solar-like stars observed by \kp, \citet{ref:silvaaguirre2017} found average uncertainties of $\sim 2\%$ in radius, $\sim 4\%$ in mass and $\sim 10\%$ in age. Clearly both the radius and the age of the star can be determined to a high level of precision, which allows for precise planetary radii to be derived while having good knowledge of the age of the system.

\section{Comparison of stellar properties from different methods}
\begin{figure}[htbp]
  \centering
    \includegraphics[width=.8\textwidth]{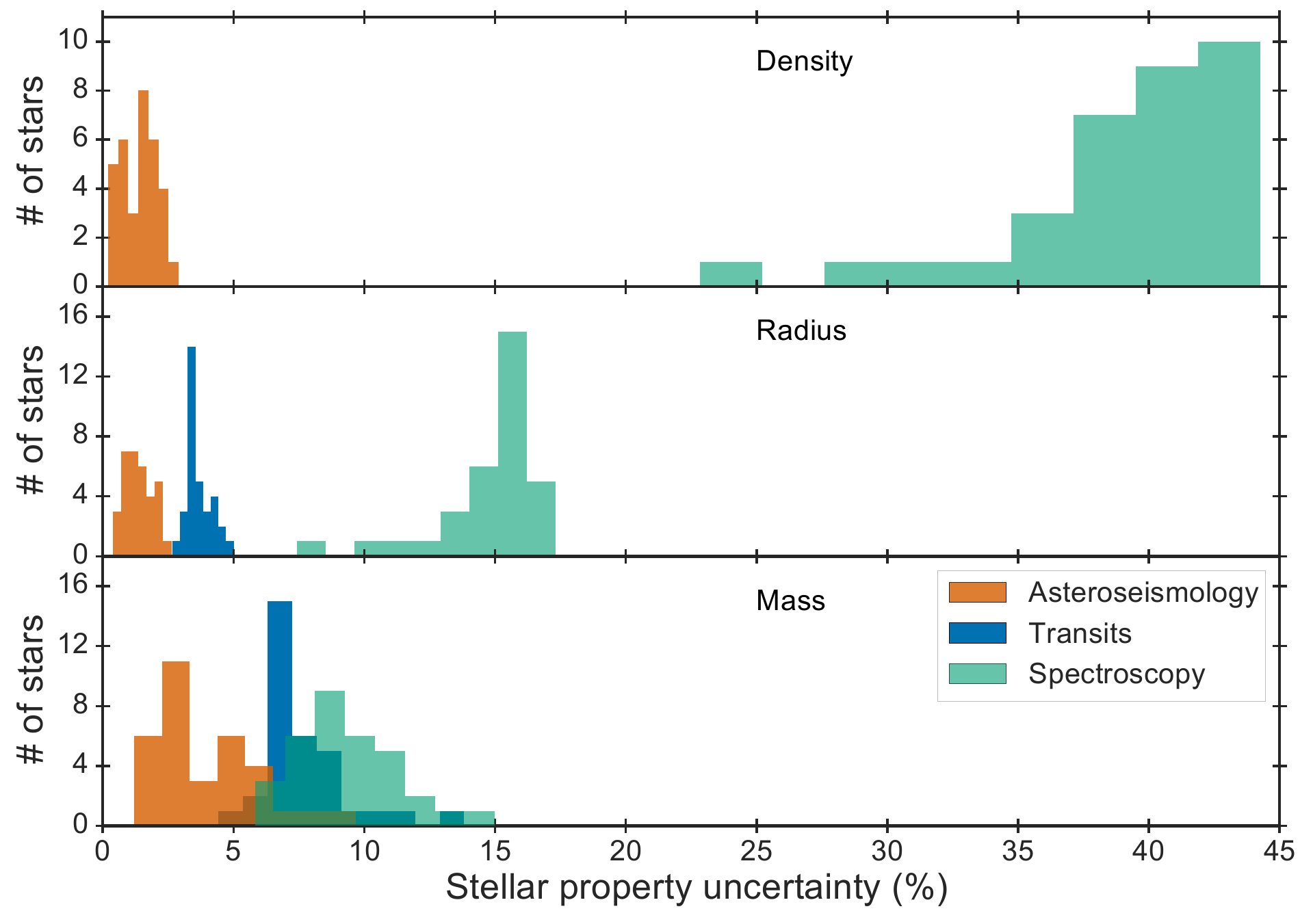}
\caption{Distribution of fractional uncertainty in stellar density (top), radius (middle) and mass (bottom) using different methods (see the legend). The sample corresponds to the exoplanet host-stars analysed in \citet{ref:silvaaguirre2015}.}
\label{fig:dens_comp}
\end{figure}

The level of precision achievable when determining stellar properties will not only depend on the uncertainties in the fitted observables, but also which observables that are being reproduced. Of particular importance for studies of planetary systems is the density of the host star, since an independent measurement of this parameter can be used in the transit fit to constrain the orbital eccentricity of an exoplanet (this is described in more detail later in this chapter). The stellar radius is also highly relevant as it, for a transiting planet, allows for the determination of the planetary radius (see Eq.~\ref{eq:dF}). Figure~\ref{fig:dens_comp} compares the obtained uncertainties in stellar density, radius and mass using asteroseismology, transits and spectroscopy for the sample of $33$ \kp\ exoplanet 
hosts studied by \citet{ref:silvaaguirre2015}.
The spectroscopic uncertainties have been computed using measured uncertainties on the temperature and metallicity and assuming a constant uncertainty on $\log g$ of $0.08$~dex. In the case of the transit uncertainties, a constant uncertainty on the density of $5\%$ was assumed in addition to what was the case for the spectroscopic sample, since this is an optimistic estimate of the density uncertainty that can be determined from the transit. The asteroseismic uncertainties have been determined from fitting individual frequencies in addition to temperature and metallicity.

The figure illustrates that modelling individual frequencies tightly constrain the fractional uncertainties; for example all density uncertainties are below 3\% and the median value is 1.7\%, which is in stark contrast to the density uncertainties derived from spectroscopy where the uncertainty is larger than $20\%$ in all cases. More generally for all the parameters shown, asteroseismology provides the highest precision, followed by transits and then spectroscopy. For the stellar mass this is the least clear-cut, owing partly to the fact that - out of the three parameters shown - the mass is the least constrained directly from the asteroseismic information.

\section{Asteroseismology of exoplanet hosts}
\begin{figure}
	\centering
	\includegraphics[scale=.4]{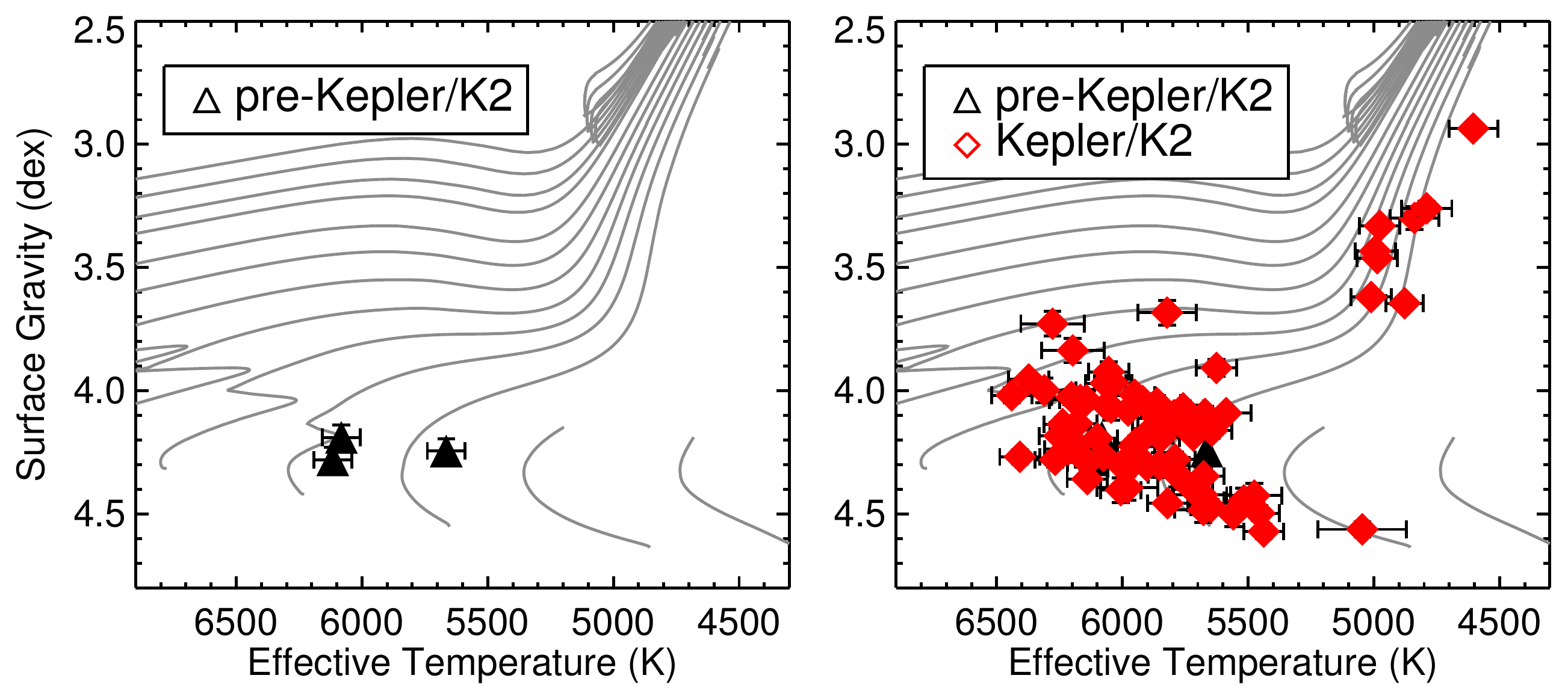}
	\caption{Surface gravity plotted against effective temperature for host stars of confirmed exoplanets with an asteroseismic detection. The left panel shows the sample before the launch of \kp/K2 ($\mu$Ara, HD17156 and HD52265) and the right panel the current sample. Grey lines are solar-metallicity evolutionary tracks from the BASTI database \citep{ref:pietrinferni2004}. From \citet{ref:huber2018}.}
	\label{fig:hrd}
\end{figure}
\noindent Especially since the launch of \kp, asteroseismology has started to play an important role for the characterisation of exoplanets and exoplanet systems. This synergy has proven very powerful and has yielded many interesting results. To date, the sample of exoplanet host stars with an asteroseismic detection exceeds 100 with \kp\ alone, if the host stars of planet candidates are also included \citep{ref:huber2013,ref:lundkvist2016}. The sample before and after the launch of \kp\ can be seen in Fig.~\ref{fig:hrd}, where it is evident that \kp\ has spearheaded a transformation in the synergy between asteroseismology and exoplanet science.

The primary contribution of asteroseismology has been to determine highly precise host star radii, which translates into very precise planetary radii when combined with the transit depth (see Eq.~\ref{eq:dF}). Some of the most precisely determined planetary radii have uncertainties as low as $1-2\%$ \citep{ref:silvaaguirre2015}, which in some cases correspond to $\sim 125$~km \citep[see for instance][]{ref:ballard2014,ref:fogtmann2014}.

Knowing the radius of a planet is important for understanding its composition, since composition models depend sensitively on this parameter. For instance, the transition from a rocky to a volatile-rich composition is estimated to take place at $R_\mathrm{p} \sim 1.6 \ R_\oplus$ \citep{weiss14,rogers15}. Uncertainties in the planetary radius can therefore cast doubt on whether a given exoplanet is rocky or gaseous. An example of this is the exoplanet Kepler-452b. It orbits a Sun-like star in the habitable zone, but has an uncertain radius just on the predicted threshold between rocky and gaseous \citep{jenkins15}. Therefore, it is very difficult to determine whether it is a rocky planet or not.

Why the stellar and planetary radii are so intimately linked can be understood by assuming that a planetary transit can be described as two spheres passing in front of each other. The transit depth ($\Delta F$) can then be found as \citep{ref:seager2003}
\begin{equation}
	\label{eq:dF}
	\Delta F = \left( \frac{R_\mathrm{p}}{R_*} \right) ^2 ,
\end{equation}
with $R_\mathrm{p}$ and $R_*$ being the planetary and stellar radius respectively. In practice, effects such as limb darkening and a non-zero impact parameter have to be taken into account. However, detailed modelling of the transit lightcurve can yield planet-to-star radius ratios that are precise enough that for $99\%$ of all planet candidates detected by \kp\ the dominating contribution to the uncertainty on the planetary radius is the uncertainty on the stellar radius \citep{ref:huber2018}. This is one of the reasons why asteroseismology is so important for the precise characterisation of exoplanets.

Another example is the Kepler-444 system. Here, using asteroseismology it has been possible to determine both precise radii and a precise age of the system, and to thereby establish that it is the oldest known system with terrestrial-sized exoplanets. The Kepler-444 system has an age of $11.2 \pm 1.0$~Gyr, which means that it formed when the Universe was about $20\%$ of its current age \citep{ref:campante2015}. The system is a compact system with five planets, which all have radii between those of Mercury and Venus and all orbit closer to their host star than Mercury does to the Sun. This results shows that small planets have formed throughout a large fraction of the history of the Universe. A pair of low-mass stellar companions in a highly eccentric orbit have also been detected in this system, which shows that the formation of small planets can also occur in a truncated protoplanetary disk \citep{ref:dupuy2016}.

\subsection{Orbital eccentricities of small exoplanets}

Knowing the eccentricity of an exoplanet's orbit is important as it can hold clues to its formation and evolution and also impact its habitability. However, determining the eccentricity of small exoplanets is difficult with traditional methods, which favour large planets \citep[Doppler velocities,][]{ref:marcy2014} and a small subset of multiplanet systems \citep[timing of secondary transits and transit timing variations, e.g.][]{ref:hadden2014}. For an overview the reader is referred to \citet{ref:hadden2017}.

If an exoplanet is on an eccentric orbit, its orbital speed will change depending on its position in the orbit according to Kepler's second law of motion. As a consequence, the observed transit duration will change depending on where in the orbit of the exoplanet the transit happens. If it occurs on a slow-moving part of the orbit, the transit duration will be longer than in the circular case, and vice versa if the transit takes place while the planet is on a fast moving portion of the orbit.

The transit duration for a transiting exoplanet moving on a circular orbit is governed by the mean stellar density \citep{ref:seager2003}. The mean stellar density assuming a circular orbit and the true stellar mean density are related as \citep[e.g.][]{ref:kipping2010}
\begin{equation}
	\label{eq:rhos_rhotr}
	\frac{\rho_*}{\rho_\mathrm{tr}} = \frac{\left( 1 - e^2 \right) ^{3/2}}{\left( 1 + e \sin \omega \right) ^3} \ ,
\end{equation}
with $e$ being the eccentricity and $\omega$ the angle of periastron. Thereby, if the mean stellar density is known from another method, such as asteroseismology, transits can be used to constrain the orbital eccentricity of the exoplanet.

\begin{figure}
	\centering
	\includegraphics[scale=.38]{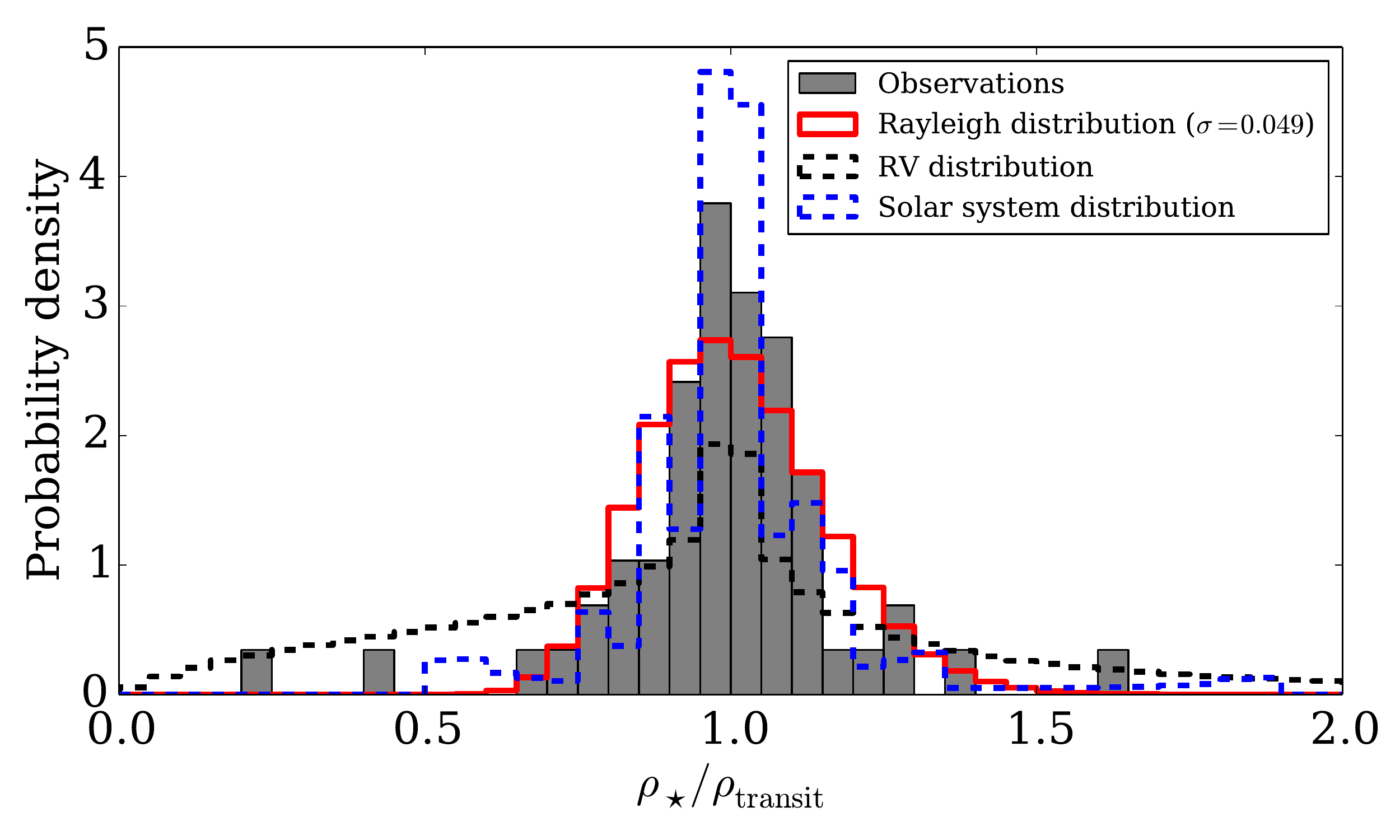}
	\caption{Ratio of the stellar mean density derived from asteroseismology and transits assuming a circular orbit for 28 \kp\ multiplanet host stars (grey). Corresponding distributions for planets detected using radial velocities (black) and the solar system planets (blue) are also shown. From \citet{ref:vaneylen2015}, \copyright AAS. Reproduced with permission.}
	\label{fig:vaneylen_ecc}
\end{figure}

First studies combining asteroseismically determined stellar densities with those measured from transits for the Kepler sample were done by \citet{ref:sliski2014} and \citet{ref:vaneylen2015}. The latter investigated a sample of 28 multiplanet systems, which are expected to have a high fraction of true planets \citep{ref:lissauer2012}. A histogram of the derived ratio between seismic and transit density for their sample of 74 exoplanets can be seen in Fig.~\ref{fig:vaneylen_ecc}, which also includes a sample of planets with eccentricities determined from radial velocities and the solar system planets, for comparison. As is the case for the solar system, the sample with asteroseismology have ratios that are close to unity and thus consistent with circular orbits, something that is different to what is seen for the radial velocity sample that displays a wider range of ratios. Given that \kp\ multiplanet systems preferentially contain small planets \citep{ref:latham2011}, this indicates that small planets tend to have circular orbits. This is an important result, not only because it may hold clues to understanding the evolution of these systems, but also because eccentricity can impact both the habitability of exoplanets and predicted occurrence rates \citep{ref:vaneylen2015}.

The sample of planets with eccentricities was expanded by \citet{ref:xie2016} and \citet{ref:hadden2017}, although not using asteroseismology. \citet{ref:xie2016} used spectroscopically determined stellar densities to confirm that \kp\ multiplanet systems indeed seem to have circular orbits, while single planet systems show larger eccentricities. That multiplanet systems have low eccentricities is in agreement with what was found by \citet{ref:hadden2017} using transit timing variations, but they add that the eccentricity is in many cases non-zero. Further expanding these studies using asteroseismology to also investigate, for instance, single planet systems would be valuable to independently confirm the results by \citet{ref:xie2016} using a high precision sample.

\subsection{Using precise planet properties to confirm evaporation}

Super-Earths and sub-Neptunes are the most common type of planet found by \kp\ \citep{ref:borucki2011,ref:batalha2013,ref:marcy2014}. In spite of this, a staggering absence of these planets in ultra short period orbits (USP planets) is evident in the \kp\ data, which has been attributed to photo-evaporative stripping of the volatile-rich envelopes of these planets. That photo-evaporation plays a role in sculpting the population of hot, close-in planets has been predicted for some years \citep[e.g.][]{ref:lopez2013,ref:owen2013}, and the observational evidence to confirm this has been provided using asteroseismology. \citet{ref:lundkvist2016} performed an asteroseismic analysis of over 100 exoplanet host stars, harbouring in excess of 150 exoplanets, and used the very precise stellar radii and mean densities obtained to compute highly precise planet properties.

This was done by combining the stellar radius and mean density from asteroseismology with the orbital period of the planet and the ratio of planet to stellar radii (see Eq.~\ref{eq:dF}). This allows for the planet radius to be found simply as the product of the stellar radius and the planet-to-star radius ratio. Including also the stellar effective temperature, the flux incident on the exoplanet from the star can be found from re-writing Kepler's third law and combining it with the inverse-square law for the flux \citep{ref:lundkvist2016}:
\begin{equation}
	\label{eq:incidentflux}
	\frac{F}{F_\oplus} =
		\left( \frac{\bar{\rho}_*}{\bar{\rho}_\odot} \right) ^{-2/3}
		\left( \frac{P}{1 \ \mathrm{yr}} \right) ^{-4/3}
		\left( \frac{T_{\mathrm{eff},*}}{T_{\mathrm{eff},\odot}} \right) ^4 \ .
\end{equation}
Here, $\bar{\rho}$ is the mean density, $P$ is the orbital period of the exoplanet, and $T_\mathrm{eff}$ is the effective temperature with the subscript $*$ indicating the star and $\odot$ the Sun.

\begin{figure}[htbp]
  \centering
  \begin{minipage}[b]{0.48\textwidth}
    \includegraphics[width=\textwidth]{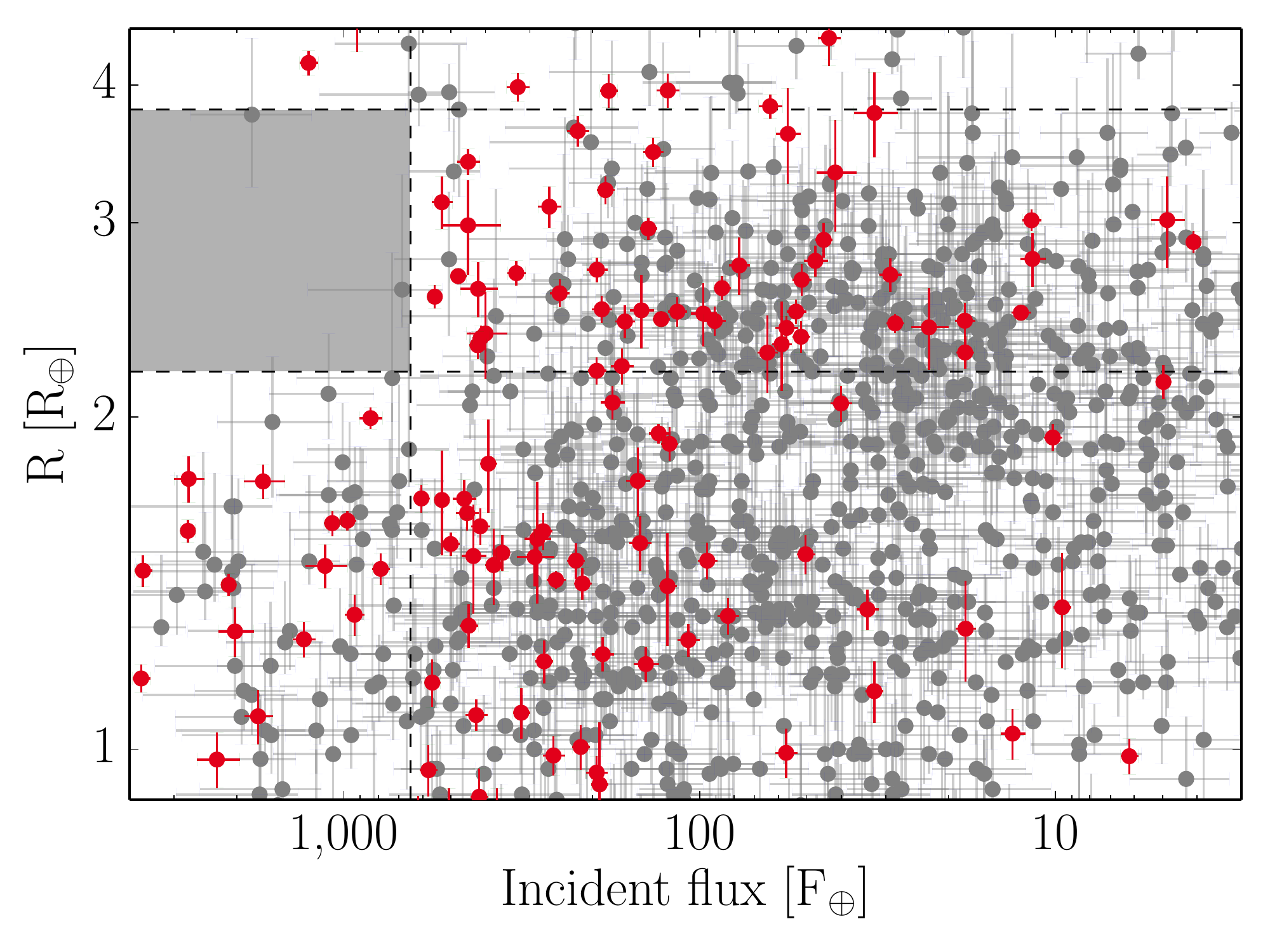}
  \end{minipage}
  \hfill
  \begin{minipage}[b]{0.48\textwidth}
     \includegraphics[width=\textwidth]{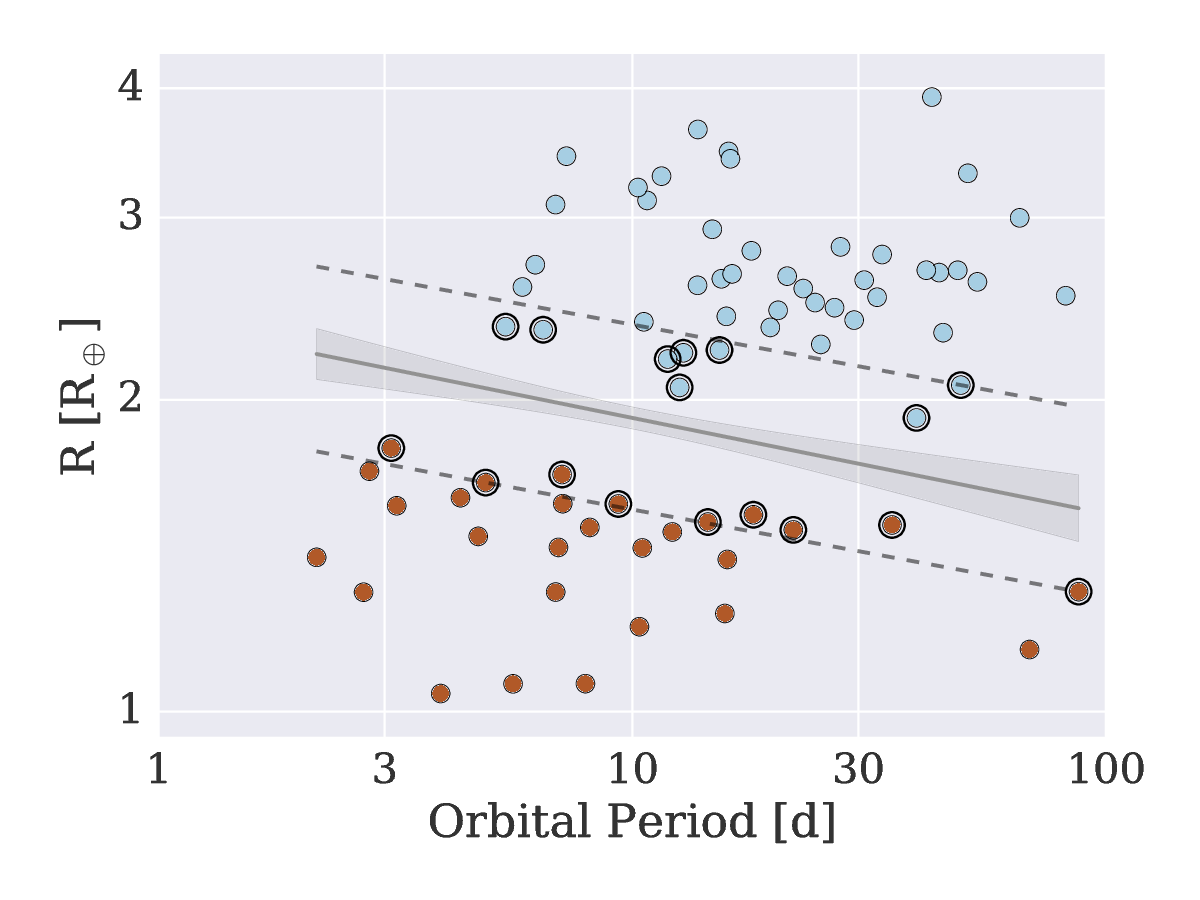}
  \end{minipage}
\caption{Two features caused by evaporation are visible in these radius-flux/period diagrams; the desert of hot super-Earths/sub-Neptunes (shaded area, left panel) and the evaporation valley (shaded area, right panel). Left: The red points highlight the planets with seismic host stars, while other \kp\ planets are shown in grey \citep[adapted from][]{ref:lundkvist2016}. Right: The seismic exoplanet sample with the planets used to constrain the slope of the valley encircled \citep[from][]{ref:vaneylen2017}.}
\label{fig:evap}
\end{figure}

A desert of hot super-Earths/sub-Neptunes can be observed for USP planets when plotting the planetary radius against either the planetary period or the incident flux. This can be seen in the left panel of Fig.~\ref{fig:evap} for the incident flux. The shaded region highlights the evaporation desert found by \citet{ref:lundkvist2016};  that is a region of parameter space void of exoplanets, which is consistent with the expected effect of photo-evaporation of volatile-rich atmospheres.

The observational confirmation of an evaporation desert emphasizes the importance of asteroseismology for obtaining precise exoplanetary properties, and additionally shows that evaporation plays an important role in shaping the exoplanet population that we see today. It can also be useful for constraining the formation of USP planets. For instance, it has been shown by \citet{ref:lopez2017} that in order to recreate such an evaporation desert, the majority of USP planets must have formed from water-poor material.

A recent breakthrough in our understanding of the radius distribution of small planets was achieved by \citet{ref:fulton2017}, using a large sample of host stars with radii derived from high-resolution spectroscopy. Importantly the spectroscopic surface gravities and radii were calibrated against asteroseismology \citep{ref:petigura2017}, highlighting the importance of asteroseismology as a fundamental benchmark for more indirect methods. \citet{ref:fulton2017} investigated small, close-in planets with periods less than $100$~days and detected a bimodality in the size distribution with few planets having a radius between $1.5-2.0 \ R_\oplus$. They also found an evaporation valley consistent with predictions by \citet{ref:owen2013} and \citet{ref:lopez2013}. This deficit of planets with a radius $\sim 1.75 \ R_\oplus$ along with clear evidence of an evaporation valley was also found by \citet{ref:vaneylen2017}. They used the asteroseismic samples of exoplanet host stars by \citet{ref:silvaaguirre2015} and \citet{ref:lundkvist2016} along with new light curve modelling to derive very precise planetary radii and orbital periods.

A plot showing the planetary radius as a function of the period can be seen in the right panel of Fig.~\ref{fig:evap}. Here, the evaporation valley and its slope are clearly visible, showing that the gap occurs at larger radii for lower orbital periods. Determining the slope of the valley is important, because, according to \citet{ref:lopez2016} it speaks to the formation mechanism of the hot, short period super-Earths. They argue that a negative slope is consistent with that expected for photo-evaporation, while being inconsistent with exclusive late formation of gas-poor rocky planets. The slope of the evaporation valley thus indicates that a substantial fraction of short-period super-Earths originate from evaporation of planets with a significant envelope. According to \citet{ref:lopez2016}, this could impact the determination of the frequency of Earth-like planets in the habitable zone ($\eta_\oplus$), which further emphasizes the importance of understanding the formation of these hot, short-period super-Earths.

\subsection{Spin-axis inclinations from rotational splittings}

The obliquity or the spin-orbit angle ($\psi$) is the angle between the normal to the orbital plane of the exoplanet and the rotation axis of the host star. It is a valuable parameter to determine because of its importance for understanding the dynamical formation history and evolution of exoplanet systems.

\begin{figure}[htbp]
  \centering
  \begin{minipage}[b]{0.48\textwidth}
    \includegraphics[width=\textwidth]{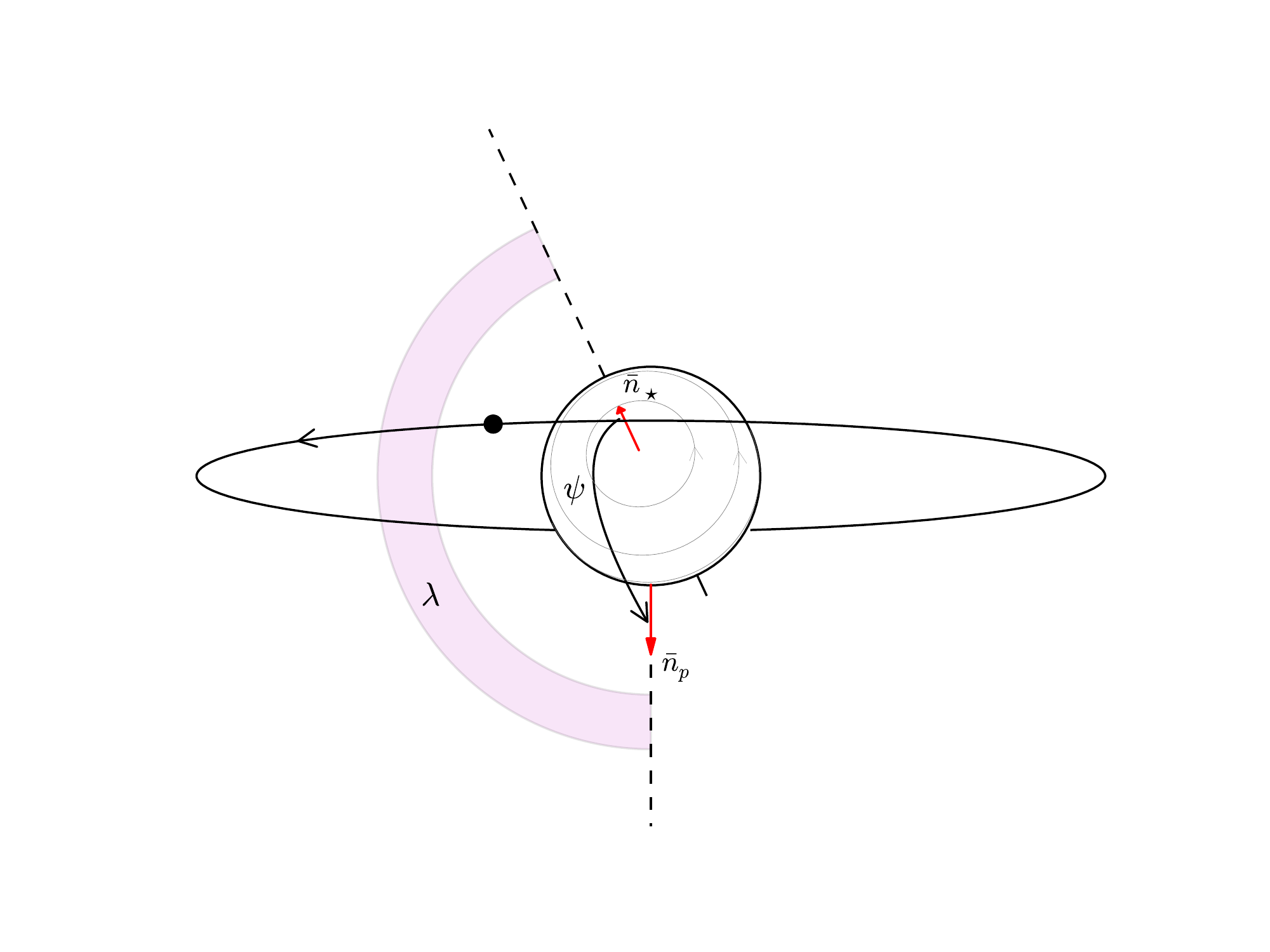}
  \end{minipage}
  \hfill
  \begin{minipage}[b]{0.48\textwidth}
    \includegraphics[width=\textwidth]{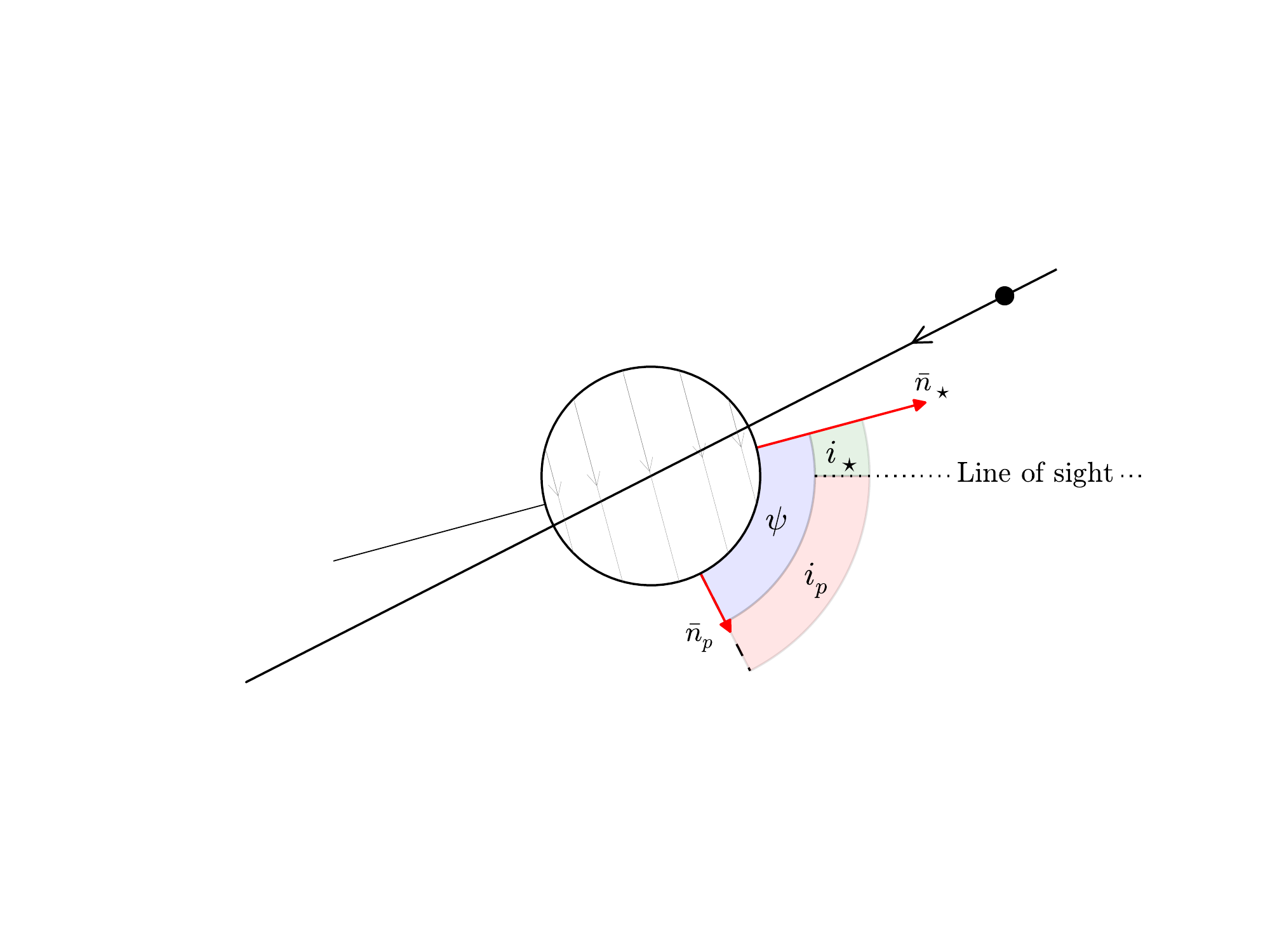}
  \end{minipage}
\caption{Illustration of the obliquity ($\psi$), the sky-projected spin-orbit angle ($\lambda$), the inclination of the orbital plane along the line of sight ($i_\mathrm{p}$) and the line of sight stellar inclination angle ($i_*$) in the HAT-P-7 system. From \citet{ref:lund2014} and reproduced with permission \copyright ESO.}
\label{fig:hatp7_angles}
\end{figure}

The obliquity may be computed as \citep{ref:fabrycky2009}
\begin{equation}
	\label{eq:obliquity}
	\cos \psi = \sin i_\mathrm{p} \cos \lambda \sin i_* + \cos i_\mathrm{p} \cos i_* \ .
\end{equation}
Here, $i_\mathrm{p}$ is the angle between the line of sight and the planetary orbital axis, $i_*$ is the inclination angle of the stellar rotation axis with respect to the line of sight, and $\lambda$ is the angle between the sky projections of the stellar spin axis and the orbital plane of the exoplanet (the sky-projected spin-orbit angle). Figure~\ref{fig:hatp7_angles} illustrates these angle in the HAT-P-7 system.

In the case of a transiting exoplanet, $i_\mathrm{p}$ can normally be determined from the transit lightcurve (it will be $\sim 90^\circ$), and $\lambda$ can be obtained through radial velocity observations during the transit (the Rossiter-McLaughlin effect). As alluded to earlier, the value of $i_*$ can be determined through asteroseismology by measuring relative amplitudes of rotationally split multiplets \citep{ref:gizon2003}. Thereby, using both information from the transit, from spectroscopic observations and from asteroseismology we can establish a full 3D view of the star-planet system.

One system for which this has been done is the HAT-P-7 system, shown schematically in Fig.~\ref{fig:hatp7_angles} \citep{ref:benomar2014,ref:lund2014}. Here, it was found from the Rossiter-McLaughlin effect that the planet, HAT-P-7b, is likely orbiting against the stellar rotation \citep{ref:winn2009,ref:narita2009,ref:albrecht2012}, but without knowing the line of sight stellar inclination angle, the orbit of the planet could be retrograde or closer to polar. From asteroseismology it was determined that the star is viewed almost pole-on \citep[$i_* < 36.5^\circ$,][]{ref:lund2014}, which indicates that the planet is most likely in a retrograde but near-polar orbit.

Often, it is not possible to determine the obliquity directly, since all the needed measurements are only available for a small number of systems. It is worth noting that for a transiting planet, if the stellar inclination angle is measured to be close to a pole-on view ($~\sim 0^\circ$), this indicates that the star-planet system is misaligned (high obliquity), since a transit shows that the orbital plane is (almost) aligned with the line of sight. However, a stellar inclination angle close to $90^\circ$ (equator-on view) does not necessarily imply an aligned system \citep[see e.g.][]{ref:campante2016}.

Estimating the obliquity in a statistical fashion has been done for a number of systems using asteroseismology, including Kepler-50 and Kepler-65 \citep{ref:chaplin2013_ob}, Kepler-410A  \citep{ref:vaneylen2014}, Kepler-432 \citep{ref:quinn2015} and 16 Cygni B \citep{ref:davies2015}. All multiplanet systems in these studies have been found to be well aligned, supporting the theory that high obliquities are confined to hot Jupiters and thus likely related to their formation \citep[e.g.]{winn10,ref:albrecht2013}.

The first counterexample of this observed trend of well-aligned multiplanet systems was also provided by asteroseismology: Kepler-56, a red giant hosting two transiting planets \citep{ref:huber2013_sci}. Here, asteroseismology yielded an inclination angle of $i_* = 47^\circ \pm 6^\circ$, which implies spin-orbit misalignment with an obliquity of $\psi > 37^\circ$ \citep{li14}. Follow-up observations have confirmed a massive, non-transiting planet in a wide orbit in this system, which is believed to be responsible for the misalignment \citep{boue14,otor16}.

A study using an asteroseismic determination of the stellar inclination angle of 25 main sequence and sub-giant solar-like stars was carried out by \citet{ref:campante2016}. They used the determined $i_*$ to place statistical constraints on the obliquity of the systems. They found that for the 25 systems in their sample, 6 of them are potentially misaligned (including HAT-P-7), but all systems are consistent with spin-orbit alignment within $2\sigma$. Therefore, Kepler-56 is a unique example of an unambiguously misaligned multiplanet system. Further obliquity measurements using asteroseismology will be important in order to determine whether spin-orbit misalignments in multiplanet systems are common or not.

\subsection{Other types of host stars}
\begin{figure}
	\centering
	\includegraphics[scale=.38]{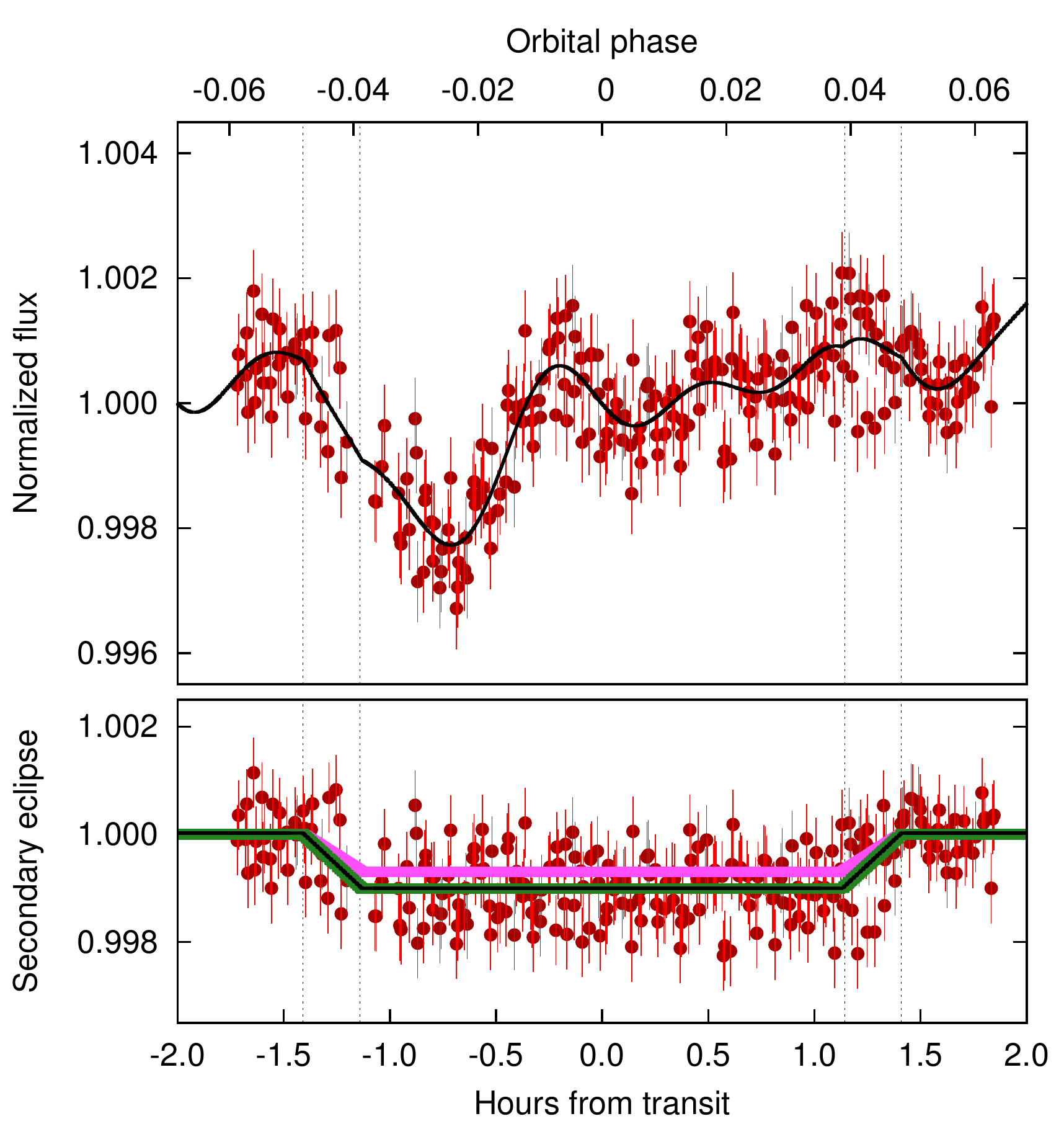}
	\caption{$\delta$~Scuti pulsations and secondary eclipse in the WASP-33 system (top). With the pulsations removed, the secondary eclipse is clear (bottom). The black lines show the best fit to the pulsations and the secondary eclipse, while the pink and green curves are two different eclipse models, and the dashed lines show where the first to fourth contacts occur. Adapted from \citet{ref:vonEssen2015} and reproduced with permission \copyright ESO.}
	\label{fig:wasp33}
\end{figure}

\noindent Asteroseismology of exoplanet hosts is not restricted to  solar-like oscillators. One example of this is the host star WASP-33, which is orbited by the hot Jupiter WASP-33b and shows $\delta$~Scuti-type pulsations. These can be found in stars in the mass range from $\sim 1.5 \ M_\odot$ to $\sim 2.5 \ M_\odot$ \citep{ref:pamyatnykh2000,ref:murphy2017} and show larger pulsation amplitudes and fewer pulsation frequencies compared to the solar-like oscillators. \citet{ref:vonEssen2015} determined the temperature of the planet WASP-33b by measuring the depth of the secondary transit. However, as is also clear from Fig.~\ref{fig:wasp33}, this was only possible after identifying the pulsation frequencies and removing them from the light curve, as the secondary eclipse was otherwise completely hidden within the signal from the stellar pulsations.

Another example could potentially come from the system with the hottest planet found to date; KELT-9 \citep[or HD195689,][]{ref:gaudi2017}. The planet KELT-9b is hotter than some stars with a temperature exceeding $4000$~K. The nominal temperature of the host star puts in between the classical instability strips for $\delta$~Scuti or slowly pulsating B stars, but future observations with improved precision may reveal pulsations since instability strip borders are not well-defined \citep{ref:handler2013}. Additional high-impact studies could come from intermediate-mass host stars such as the $\gamma$\,Doradus pulsator HR8799 \citep{ref:zerbi99}, for which space-based asteroseismology by missions such as TESS will dramatically improve our understanding of the age of the star and its directly imaged planets \citep[e.g.][]{ref:marois10}.

Mode identiﬁcation in classical pulsators such as $\delta$~Scuti stars is challenging; however the extension of the asteroseismology-exoplanet synergy to these systems will without a doubt become more important in the near future. As an example, the photometric observations that will be provided over several years by the PLATO mission may give future opportunities to discover planets in wide orbits around $\delta$~Scuti stars by exploiting the small shifts in the pulsation frequencies induced by the planet. This method has already been used by \citet{ref:murphy2016} to detect a $12 \ M_\mathrm{Jup}$ planet in an $840$~day orbit around its $\delta$~Scuti host star. At this orbital distance the planet is in or near the habitable zone, making it the first planet discovered within $1\sigma$ of the habitable zone around this type of star.

\begin{acknowledgement}
The authors would like to thank Vincent Van Eylen, Carolina von Essen and Mikkel Lund for providing figures for this manuscript. Funding for the Stellar Astrophysics Centre is provided by The Danish National Research Foundation (Grant DNRF106). M.S.L. is supported by The Independent Research Fund Denmark's Sapere Aude program (Grant agreement no.: DFF–5051-00130). D.H. acknowledges support by the National Aeronautics and Space Administration under Grant NNX14AB92G issued through the Kepler Participating Scientist Program. V.S.A. acknowledges support from the Villum Foundation (Research grant 10118). W.J.C. acknowledges support from the UK Science, Technology and Facilities Council (STFC).
\end{acknowledgement}

\bibliographystyle{spbasicHBexo}  
\bibliography{lundkvist}

\end{document}